\documentstyle[epsf,aps]{revtex}

\begin{document}
\title{ Medium modifications of the nucleon--deuteron break--up
  cross section in the Faddeev approach}
\author{M. Beyer, G. R\"opke and A. Sedrakian\footnote{
on leave of absence from Physics Department, Yerevan State University,
Armenia}} 
\address{Max--Planck--Gesellschaft, AG 
`Theoretische Vielteilchenphysik'\\
 Universit\"at Rostock, Universit\"atsplatz 1, 
18055 Rostock, Germany}

\maketitle

\begin{abstract}
  The three--nucleon scattering problem in a nuclear medium is
  considered within the Faddeev technique.  In particular the deuteron
  break--up cross section that governs the formation and the break--up
  reactions of deuterons ($NNN\leftrightarrow Nd$) in a nuclear
  environment is  calculated at finite temperatures and densities. A
  significant enhancement of the in--medium break--up cross section with
  increasing density has been found.
\end{abstract}

\section*{}
Formation of light clusters such as deuterons, helium and alpha
particles is an important aspect of heavy ion collisions at
intermediate energies, see e.g.~\cite{nag81}. Empirical evidence,
including recent experimental data on cluster formation, see
refs.~\cite{GSI93,MSU95}, indicate that a large fraction of deuterons
can be formed in heavy--ion collision of energies $E/A\le 200$ MeV.

During the expansion of the system the density can drop below the
Mott--density of deuteron dissociation~\cite{bal95,sch90,alm90}.  
In this region the deuteron abundances will be determined by deuteron
formation, $NNN\rightarrow dN$, and break--up, $dN\rightarrow NNN$,
reactions.  Since the deuteron formation rate can be expressed through
the break--up probability, the final outcome of the reaction will be
essentially controlled by the deuteron break--up cross section.
Previous studies of the kinetics of deuteron production have utilized
the impulse approximation to calculate the reaction cross section at
energies above 200 MeV/A~\cite{dan91}. For lower energies, viz. $E/A
\le 200$ MeV, the impulse approximation fails and a full three--body
treatment of the scattering problem is necessary.  Furthermore, a
consistent treatment of cluster formation in expanding hot and dense
matter requires the inclusion of medium effects into the respective
reaction cross sections.

The essentials of the three--body problem in the vacuum case are well
known, see e.g. ref.~\cite{glo88}.  In the following we utilize the
AGS formalism~\cite{alt67} suitably modified, in order to treat the
three--body problem taking into account the nuclear medium in the
quasi--particle approximation. To this end, we will rely on a separable
representation of the nucleon--nucleon potential.  This choice
simplifies the problem considerably.  A systematic investigation of
separable parameterizations of ``realistic'' potentials has been
pursued e.g. by Plessas and collaborators~\cite{ple95}.  We note that
solutions of the three--body problem using ``realistic'' $NN$
potentials have been achieved e.g. by the Bochum group~\cite{glo90},
and the Bonn group in the framework of the $W$--matrix
approach~\cite{san90}.

Recently, separable approximations have been used to solve the pion
deuteron scattering problem~\cite{hui90D} and successfully applied to
coherent photo-- and electro--production of pions on the
deuteron~\cite{bla95}. A detailed discussion of the numerical method
used in these cases has been given in~\cite{hui90}. In the present
work we have modified this pion deuteron code~\cite{ben95} to solve
the nucleon--deuteron scattering problem. Our results for the dominant
three--body transition matrix elements agree within $\simeq 5\%$ with
the previous calculations by Doleschall \cite{dol73}.

The AGS equations for the transition matrix
$U_{\alpha\beta}(z)$ are  given in a compact notation by~\cite{alt67}
\begin{equation}
U_{\beta\alpha}(z)=(1-\delta_{\alpha\beta}) G_0^{-1}(z) 
+\sum_{\gamma=1}^3 (1-\delta_{\alpha\gamma}) 
T_\gamma(z) G_{0}(z) U_{\gamma\alpha}(z),
\label{ags}
\end{equation}
where the Greek indices represent particle channels.  In the
absence of the medium the free three--body
resolvent is  $G_0(z) = (z-H_0)^{-1}$ with $ z=E+i\varepsilon $,
where $H_0$ is the Hamiltonian of the noninteracting three--body system
and $E$ is the three--body energy.
The three--body transition matrix $T_\gamma(z)$ in the channel $\gamma $ is
related to the $t$--matrix of the two--body subspace $\hat T_\gamma(\hat
z)$ (denoted by ``hat'') via,
\begin{equation}
\langle \bar {\bf q}_\gamma \bar {\bf p}_\gamma| T_\gamma(z)
|{\bf q}_\gamma {\bf p}_\gamma\rangle =
\langle \bar {\bf q}_\gamma| {\bf q}_\gamma\rangle
\langle  \bar {\bf p}_\gamma|\hat T_\gamma(\hat z_\gamma)| {\bf
  p}_\gamma\rangle, 
\label{twot}
\end{equation}
with $\hat z_\gamma = z - 3 q_\gamma^2/(4m)$, where $m$ is the nucleon mass,
and ${\bf p}_\gamma $ (${\bf q}_\gamma$) are the three--body momenta in
the center of mass frame for the pair (odd) nucleon, respectively.
For a  separable two--body potential, which is given by
(overall particle channel index $\gamma $ suppressed)
\begin{equation}
\hat V_r= |g_{r}\rangle\lambda_{r}\langle g_{r}|,
\end{equation}
where $r$ denotes a set of quantum numbers providing a complete
specification of the two particle state, the two--body $t$--matrix is
then given by
\begin{equation}
\hat T_r(\hat z)= |g_{r}\rangle
\hat\tau_{r}(\hat z)
\langle g_{r}|.
\label{twotsep}
\end{equation}
Solving the two particle Lippmann--Schwinger equation determines
$\hat\tau_r$ (see e.g.~\cite{sch90})
\begin{equation}
\hat\tau_{r}(\hat z)
=\left(\lambda_{r}^{-1} - 
\langle g_{r}|\hat G_0(\hat z)|g_{r}\rangle
\right)^{-1}.
\label{tausep}
\end{equation}

Multiplying the AGS equations with $\langle g_{\beta m}|G_0(z)$ 
[$G_0(z)|g_{\alpha m}\rangle$] from left [right] they  reduce to~\cite{lov64}
\begin{equation}
X_{{\beta n,}\alpha m}(z) = Z_{\beta n,\alpha m}(z)
+ \sum_{\gamma r} Z_{\beta n,\gamma r}(z)\tau_{\gamma r}(z)
X_{\gamma r,\alpha m}(z),
\label{lovelace}
\end{equation}
where
\begin{eqnarray}
X_{\beta n,\alpha m} &= &
\langle g_{\beta n}|G_0(z)U_{\beta\alpha}G_0(z)| g_{\alpha m}\rangle,\\
Z_{\beta n,\alpha m} &= &
\langle g_{\beta n}|(1-\delta_{\beta\alpha})G_0(z)| g_{\alpha
  m}\rangle.
\label{lovedef}
\end{eqnarray}

\begin{figure}[p]
  \leavevmode
  \epsfxsize=0.80\textwidth
  \epsffile{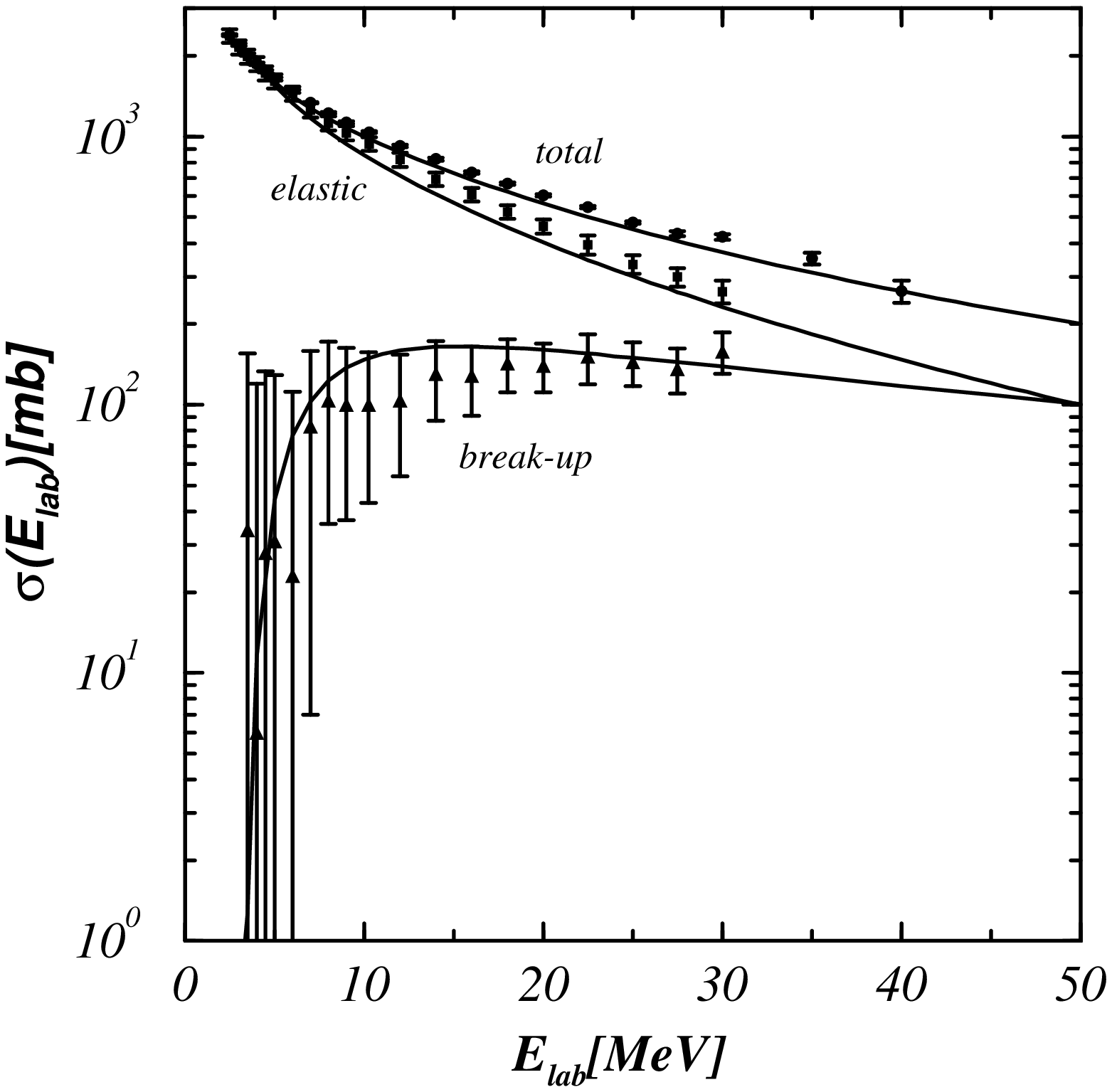}
\caption{\label{expt} A comparison of the total, elastic, 
and break--up cross sections
  with the experimental data of Ref.~[21].} 
\end{figure}

We restrict the two--body channels to the dominant ones, i.e.  $^1S_0$
and $^3S_1- {^3D}_1 $. For the separable ansatz we use the
parameterization of Phillips~\cite{phi68}. The parameters are taken
from Ref.~\cite{bru77}, which shows that gross features of the elastic
and break--up cross sections as well as the differential elastic cross
section up to $E_{lab}=50$ MeV are sufficiently well reproduced. To
calculate the break--up cross section we use the optical theorem.
Figure~\ref{expt} shows our results for the integrated cross section
along with the experimental data~\cite{sch83}.

Note that the agreement between the present calculations on free
nucleon--deuteron scattering and the experimental data can still be
improved, through using more realistic potentials and including
contributions of higher partial waves.  However, our present goal is
to study the general features related to the inclusion of nuclear
medium, which are expected to be quite significant, cf.~\cite{alm95}.

The influence of the nuclear medium on the three particle problem is
treated in mean--field approximation. Here, two effects appear, i) self
energy corrections and ii) Pauli blocking. The single particle
propagator is obtained within the Hartree--Fock approximation of the
self energy. The Hartree--Fock energy shift is given by 
($V(12,12)\equiv \langle 12|\hat V|12\rangle$)
\begin{equation}
\Delta^{HF}(1) = \sum_{2}
\left[ V(12,12)-  V(12,21) \right]\,  f(2),
\end{equation}
where the multi--indices $1,2$ denote momentum and other quantum
numbers of the respective particles, the Fermi function is given by
$f(\varepsilon_1)=\left\{\exp[(\varepsilon_1 - \mu)/T]+1\right\}^{-1}
$, and the single particle energy by $\varepsilon_1= k_1^2/2m +
\Delta^{HF}(1)$. The Hartree--Fock shift is evaluated using the same
potential as for solving the AGS equations. To simplify the three--body
treatment we introduce the effective mass approximation. The
quasi--particle dispersion relation is then written as
$\varepsilon_1=k_1^2/2m^* + \Delta_0^{HF}$, where the constant term
$\Delta_0^{HF}$ can be absorbed into the chemical potential $\mu $.
The effective mass $m^*$ is density and temperature dependent, and
decreases from the value $m^*/m=1$ in the zero density limit to e.g.
$m^*/m=0.97$ at $T=10$ MeV for the Mott density of the deuteron
$n=8\times 10^{-3}$ fm$^{-3}$.

The Bethe--Salpeter 
equation for the three--body Green's function $G(\bar 1\bar 2\bar 3,
123;z)$ in mean field approximaton, 
including Pauli blocking and self energy shifts, is obtain from the
Matsubara Green's function technique as
\begin{eqnarray}
\lefteqn{G(\bar 1\bar 2\bar 3, 123;z)
= \frac{1-f(1)-f(2)-f(3)}
{z-\varepsilon_1-\varepsilon_2-\varepsilon_3}
\;\;\langle \bar 1\bar 2\bar 3| 123\rangle}\nonumber\\
&&+\sum_{1'2'3'}G(\bar 1\bar 2\bar 3, 1'2'3';z)
\left(V( 1'2',12)\;\langle 3'|3\rangle\;\;
\frac{1-f(1)-f(2)}
{z-\varepsilon_1-\varepsilon_2-\varepsilon_3}
 + \mbox{cycl. perm.}\right)
\end{eqnarray}
This leads to modifications of the quantities entering in the AGS
equations. Pauli blocking then appears in the transition matrix
$T_\gamma(z)$ and the resolvent $G_0(z)$.  The thermodynamic
$t$--matrix acting in channel $\gamma $, see eq.~(\ref{twot}), has been
derived within the Matsubara--Green's function techniques, see Ref.~\cite{alm94}.
It turns out that, in the low density approximation used here, the
effect of the Pauli blocking may be absorbed into an effective
potential. Inclusion of all medium corrections effectively results in
the following replacements in the eqs.~(\ref{twotsep}),
(\ref{tausep}), (\ref{lovelace}), and (\ref{lovedef}), viz.
\begin{eqnarray}
m &\rightarrow &m^*,\label{repm}\\
g_{\alpha m} &\rightarrow & 
g^*_{\alpha m} =  
\langle Q_\alpha \rangle^{1/2} g_{\alpha m}.
\label{repg}
\end{eqnarray}
Here $\langle Q\rangle $ is the angular averaged Pauli blocking
factor, 
\begin{equation}
\langle Q ({\bf p},{\bf q})\rangle = 
2\pi\int_{-1}^1 dx [1 - f(\varepsilon_{{\bf q}/2 + {\bf p}})
 - f(\varepsilon_{{\bf q}/2 - {\bf p}})].
\end{equation}
where $x={\bf p}\cdot{\bf q}/(pq)$.  The center of mass of the three
nucleon system is set equal to the momentum of the medium. The AGS
equations are properly symmetrized for identical particles and then
solved in the rest system of the medium.

We present our results with amplitudes normalized
to the residue of the two--body $t$--matrix in the deuteron channel $d$, i.e.
\begin{equation}
N^{*2}_{\gamma d}\equiv\mbox{Res}(\hat\tau,\hat z=E^{*B}_{\gamma d})=
\langle g^*_{\gamma d}|\hat G^2_0({E^{*B}_{\gamma d}} )
| g^*_{\gamma d}\rangle^{-1},
\end{equation}
where $E^{*B}_{\gamma d}$ is the  in--medium deuteron binding energy.
The advantage is that the optical theorem, which is used to calculate
the break--up cross section, has the same formal structure as in the
vacuum case and is valid without any further modifications. For an
alternative formulation of the in--medium optical theorem in the 
two--body case see e.g. Ref.~\cite{sch90}. The in--medium break up cross
section $\sigma_{n,T}^*(E_{lab})$ that depends on density $n$ and
temperature $T$ is then defined using the amplitudes $N^{*}_{\beta
  d} X^*_{\beta d,\alpha d} (z)N^{*}_{\alpha d}$ resulting from
solving eq.~(\ref{lovelace}) with the replacements given in 
eqs.~(\ref{repm}) and (\ref{repg}) in the same way as in the vacuum
case, see e.g.~\cite{glo88} or~\cite{hui90D,koi83}.

\begin{figure}[p]
  \leavevmode
  \epsfxsize=0.80\textwidth
  \epsffile{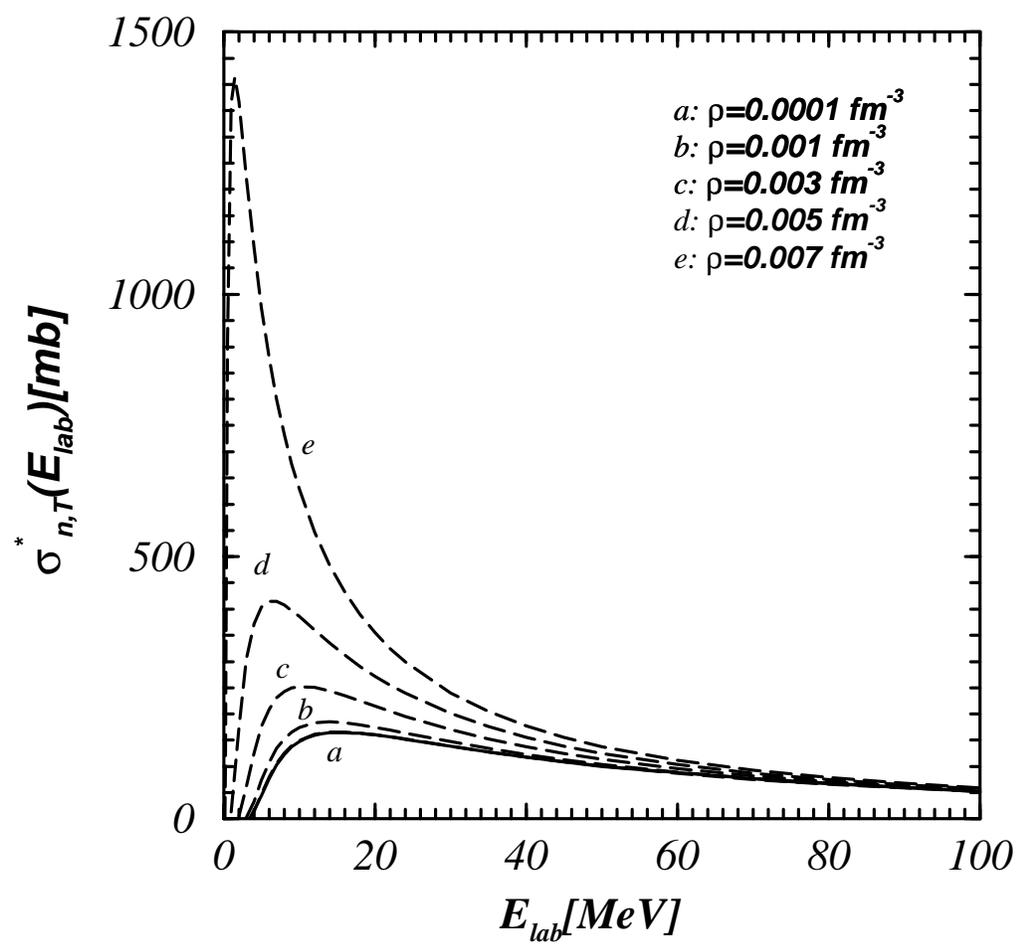}
\caption{\label{10mev} Break--up cross section at temperature $T=10$
  MeV. Free cross section is shown as solid line. Other lines are due
  to different nuclear densities, see text.}
\end{figure}

Our results are depicted in Figure~\ref{10mev}.  The solid lines
represent the free space break--up cross section as shown in
Figure~\ref{expt}. The dashed lines show the break--up cross section,
$\sigma_{n,T}^*(E_{lab})$ for densities $n=0.1,\, 1,\, 3,\, 5,\, 7\, \times
10^{-3}$ fm$^{-3}$, respectively, as a function of the laboratory
energy $E_{lab}$. The Mott transition occurs at the
density $n=8\times 10^{-3}$ fm$^{-3}$.

\begin{figure}[p]
  \leavevmode
  \epsfxsize=0.80\textwidth
  \epsffile{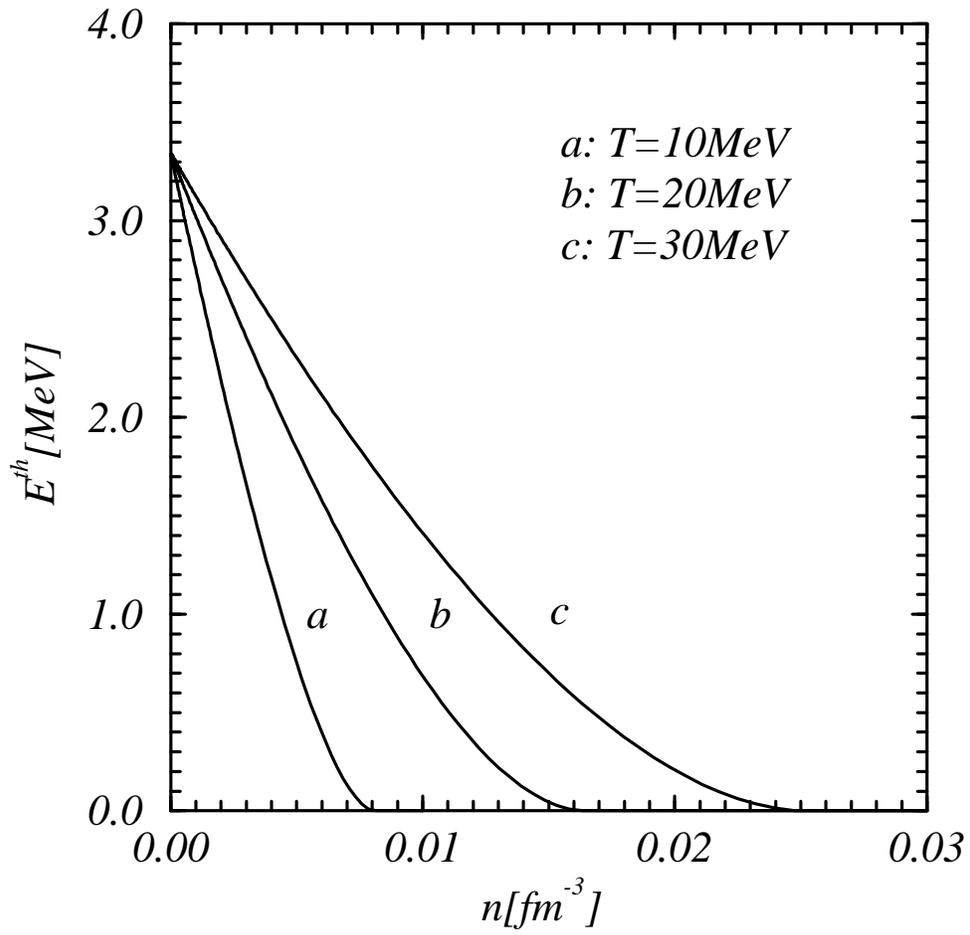}
\caption{\label{Ethresh} Shift of threshold energy $E^{th}$ in the
  laboratory system due to nuclear density $n$ at different
  temperatures $T$.}
\end{figure}

The medium effects significantly modify the vacuum break--up cross
section.  Two qualitative features are observed. First, the break--up
threshold is shifted towards lower scattering energies with increasing
density of the nuclear matter.  This kinematical effect is due to the
decrease of the deuteron binding energy with increasing density. For
an illustration of the dependence of the threshold energy in the
laboratory system $E^{th}$ on different temperatures and densities see
Figure~\ref{Ethresh}.  Second, the cross section increases
considerably with increasing density. The maximum is enhanced by one
order of magnitude for the largest density value considered.  For
densities larger than the Mott densities the deuteron disappears as a
bound state.

Also, it is instructive to see how the medium dependent cross section
converges to the ``free'' cross section.  At 100 MeV the deviation of
the in--medium cross section from the free in this model is in the
order of 10\%.  From inspection of Figure~\ref{10mev} we conclude that
the dominant changes in the cross section takes place at rather
moderate energies, i.e. where the impulse approximation fails, and the
Faddeev technique has to be used.

To summarize, we have calculated the deuteron break--up
cross section, $\sigma_{n,T}^*(E_{lab})$, 
using the AGS formalism modified in order to take into account the
effects of quasi--particle energies and the phase space occupation
effects.  It has been shown that there are strong modifications of the
deuteron break--up cross section compared to the vacuum scattering
case. A study of the reactions dynamics within a kinetic approach to
the cluster formation~\cite{tsa89,xu90,roe87}, which will
include the deuteron break--up and formation probabilities based on the
present work, would provide an improved description of the physics of
deuteron production in heavy--ion reactions.

We gratefully acknowledge B.L.G. Bakker for providing us with a
three--body pion deuteron code. One of us (M.B.) would like to thank B.L.G.
Bakker and F. Blaazer of VU Amsterdam for very fruitful discussions 
on numerical issues. G.R. and A.S. are thankful to V.B. Belyaev and
 W. Plessas for discussions.



\end{document}